\documentstyle[preprint]{jpsj}
\catcode`\@=11
\newcount\@arga
\newcount\@argb
\newcount\@argc
\newcount\@ctmpa
\newcount\@ctmpb
\newcount\@ctmpc
\newcount\@ctmpd
\newcount\@ctmpe
\newdimen\@darg
\newdimen\@bblen
\newif\if@bbllx
\newif\if@bblly
\newif\if@bburx
\newif\if@bbury
\newif\if@height
\newif\if@width
\newif\if@scale
\newif\ifno@bb
\newif\ifepsfdraft
\epsfdraftfalse
\def\@setpsfile#1{
                \typeout{epsf:[#1]}
                \def\@psfile{#1}
}
\def\@setpsheight#1{
                \@heighttrue
                \@darg=#1\relax
                \edef\@psheight{\number\@darg}
}
\def\@setpswidth#1{
                \@widthtrue
                \@darg=#1\relax
                \edef\@pswidth{\number\@darg}
}
\def\@setpsscale#1{
                \@scaletrue
                \def\@pshscale{#1}
                \def\@psvscale{#1}
                \@bblen=#1pt\relax
                \@bblen=1000\@bblen
                \def\@texhscale{\expandafter\remove@dim\the\@bblen}
                \let\@texvscale=\@texhscale
}
\def\@setpshscale#1{
                \@scaletrue
                \def\@pshscale{#1}
                \@bblen=#1pt\relax
                \@bblen=1000\@bblen
                \def\@texhscale{\expandafter\remove@dim\the\@bblen}
}

\def\@setpsvscale#1{
                \@scaletrue
                \def\@psvscale{#1}
                \@bblen=#1pt\relax
                \@bblen=1000\@bblen
                \def\@texvscale{\expandafter\remove@dim\the\@bblen}
}

\def\@setparms#1=#2,{\@nameuse{@setps#1}{#2}}
\def\ps@init@parms{
                \@heightfalse \@widthfalse
                \no@bbfalse
                \def\@psbbllx{}\def\@psbblly{}
                \def\@psbburx{}\def\@psbbury{}
                \def\@psheight{}\def\@pswidth{}
                \def\@pshscale{1}\def\@psvscale{1}
                \def\@texhscale{1000}\def\@texvscale{1000}
                \def\@psfile{}
                \def\@sc{}
}
\def\parse@ps@parms#1{
                \@for\@epsfile:=#1\do
                   {\expandafter\@setparms\@epsfile,}}
\newif\ifnot@eof
\newread\ps@stream
\def\bb@search{
        \openin\ps@stream=\@psfile
        \no@bbtrue
        \not@eoftrue
        \catcode`\%=12\relax
        \ifeof\ps@stream\typeout{epsf: File not found}\fi
        \loop
                \read\ps@stream to \line@in
                \global\toks200=\expandafter{\line@in}\relax
                \ifeof\ps@stream \not@eoffalse \fi
                \@bbtest{\toks200}\relax
                \if@bbmatch\not@eoffalse\expandafter\bb@cull\the\toks200\fi
        \ifnot@eof \repeat
        \catcode`\%=14
}       

\catcode`\%=12
\newif\if@bbmatch
\def\@bbtest#1{\expandafter\@a@\the#1
\long\def\@a@#1
        \ifx\@bbtest#2\@bbmatchfalse\else\@bbmatchtrue\fi}
\def\bb@cull 
        \@ifnextchar\space{\@latexbug}{\bb@extract}}
\def\bb@extract #1 #2 #3 #4 {
        \message{BoundingBox: (#1bp,#2bp)--(#3bp,#4bp)}
        \@darg=#1 bp\edef\@psbbllx{\number\@darg}
        \@darg=#2 bp\edef\@psbblly{\number\@darg}
        \@darg=#3 bp\edef\@psbburx{\number\@darg}
        \@darg=#4 bp\edef\@psbbury{\number\@darg}
        \no@bbfalse
}
\catcode`\%=14

\def\compute@bb{
                \bb@search
                \ifno@bb \typeout{epsf: No BoundingBox}
                \stop
                \else
                \@arga=\@psbburx
                \advance\@arga by -\@psbbllx
                \edef\@bbw{\number\@arga}
                \@arga=\@psbbury
                \advance\@arga by -\@psbblly
                \edef\@bbh{\number\@arga}
                \fi
}
\def\in@hundreds#1#2#3{\@argb=#2 \@argc=#3
                     \@ctmpa=\@argb     
                     \divide\@ctmpa by \@argc
                     \@ctmpb=\@ctmpa
                     \multiply\@ctmpb by \@argc
                     \advance\@argb by -\@ctmpb
                     \multiply\@argb by 10
                     \@ctmpb=\@argb     
                     \divide\@ctmpb by \@argc
                     \@ctmpc=\@ctmpb
                     \multiply\@ctmpc by \@argc
                     \advance\@argb by -\@ctmpc
                     \multiply\@argb by 10
                     \@ctmpc=\@argb     
                     \divide\@ctmpc by \@argc
                     \@arga=#1\@ctmpe=0
                     \@ctmpd=\@arga
                        \multiply\@ctmpd by \@ctmpa
                        \advance\@ctmpe by \@ctmpd
                     \@ctmpd=\@arga
                        \divide\@ctmpd by 10
                        \multiply\@ctmpd by \@ctmpb
                        \advance\@ctmpe by \@ctmpd
                     \@ctmpd=\@arga
                        \divide\@ctmpd by 100
                        \multiply\@ctmpd by \@ctmpc
                        \advance\@ctmpe by \@ctmpd
                     \edef\@result{\number\@ctmpe}
}
\def\compute@wfromh{
                \in@hundreds{\@psheight}{\@bbw}{\@bbh}
                \edef\@pswidth{\@result}
}
\def\compute@hfromw{
                \in@hundreds{\@pswidth}{\@bbh}{\@bbw}
                \edef\@psheight{\@result}
}
\def\compute@handw{
        \if@height 
                \if@width
                \else
                        \compute@wfromh
                \fi
        \else 
                \if@width
                        \compute@hfromw
                \else
                        \if@scale
                                \in@hundreds{\@texvscale}{\@bbh}{1000}
                                \let\@bbh=\@result
                                \in@hundreds{\@texhscale}{\@bbw}{1000}
                                \let\@bbw=\@result
                        \fi
                                \edef\@psheight{\@bbh}
                                \edef\@pswidth{\@bbw}
                \fi
        \fi
}
{\catcode`\p=12\catcode`\t=12
\gdef\remove@dim#1.#2pt{#1}}

\def\compute@sizes{
        \compute@bb
        \compute@handw
}
\def\epsfile#1{
        \ps@init@parms
        \parse@ps@parms{#1}
        \compute@sizes
        \@arga=\@psheight
        \divide\@arga by 65536
        \edef\@psvsize{\number\@arga}
        \@arga=\@pswidth
        \divide\@arga by 65536
        \edef\@pshsize{\number\@arga}
        \message{=>(\@pshsize bp,\@psvsize bp)}
        \leavevmode
        \vbox to \@psheight true sp{
                \hbox to \@pswidth true sp{
                \ifepsfdraft\hss\@psfile\hss\else
                \if@height 
                        \if@width
                                \special{epsfile=\@psfile \space 
                                hsize=\@pshsize \space
                                vsize=\@psvsize \space}
                        \else
                                \special{epsfile=\@psfile \space 
                                vsize=\@psvsize \space}
                        \fi
                \else 
                        \if@width
                                \special{epsfile=\@psfile \space 
                                hsize=\@pshsize \space}
                        \else
                                \if@scale
                                        \special{epsfile=\@psfile \space
                                        vscale=\@psvscale \space
                                        hscale=\@pshscale \space}
                                \else
                                        \special{epsfile=\@psfile \space}
                                \fi
                        \fi
                \fi
                \hfil\fi
                }
        \vfil
        }
}

\catcode`\@=12
\newcommand{\eps}{\varepsilon}
\newcommand{\vphi}{\varphi}
\def\runtitle{Vortices and Quantum tunneling 
in Current-Biased 0-$\pi$-0 Junctions}
\def\runauthor{Takeo {\sc Kato} and Masatoshi {\sc Imada}}
\title
{Vortices and Quantum tunneling 
in Current-Biased 0-$\pi$-0 Josephson Junctions
of $d$-wave Superconductors}

\author
{Takeo {\sc Kato} and Masatoshi {\sc Imada} }

\inst
{Institute for Solid State Physics, 
University of Tokyo,\\ 7-22-1 Roppongi, Minato-ku, Tokyo 106}

\recdate
{January 14, 1997}

\abst{
We study a current-biased 0-$\pi$-0 Josephson 
junction made by high-$T_{\rm c}$ superconductors,
theoretically. When a length of the $\pi$ junction is large
enough, this junction contains a vortex-antivortex pair
at both ends of the $\pi$ junction. 
Magnetic flux carried by the vortices 
is calculated using the sine-Gordon equation.
The result shows that
the magnetic flux of the vortices is suppressed to zero
as the distance between the vortices is reduced. 
By applying an external current,
the orientation of the vortices is reversed, and 
a voltage pulse is generated.
The current needed for this transition
and generated pulse energy are calculated.
Macroscopic quantum tunneling (MQT) in this transition
is also studied. The tunneling rate has
been evaluated by an effective Hamiltonian  
with one degree of freedom.}

\kword
{high-${\rm T}_{\rm c}$ superconductor,
$\pi$ junction, sine-Gordon equation, 
half vortex, macroscopic quantum tunneling}
\begin{document}
\sloppy
\maketitle
%
%
\section{Introduction}
Static and dynamical properties of long Josephson 
junctions (LJJs) has been studied by many authors
for practical applications. 
Particularly, kink propagation in LJJs has been studied
both theoretically and experimentally
on the basis of the classical (damped) sine-Gordon 
equation~\cite{Pedersen,Lomdahl}
\begin{equation}
\phi_{xx} - \phi_{tt} - \sin \phi -\alpha
\phi_t + \beta \phi_{xxt} = f .
\label{dampedsineGordon}
\end{equation}
Here, $\alpha$ and $\beta$ are damping parameters due to
the quasiparticle tunneling loss and the surface loss,
and $f=I/I_0$ is an external current scaled by the critical
current $I_0$, and the spatial variable $x$ is normalized by
the Josephson penetration length $\lambda_{\rm J}$.
Recently, quantum effects of a kink have been discussed through
interference effects,~\cite{Hermon}
and macroscopic quantum tunneling (MQT).~\cite{Kato}

In this paper, we consider a long Josephson junction
including a so-called $\pi$ junction, which has
negative critical current. 
In such junctions, 
the phase difference shows more complicated properties 
than traditional LJJs. Hence, we expect
new phenomena characteristic of the sine-Gordon field
with infinite degrees of freedom.

The negative critical current has been discussed first 
in Josephson junctions, which have an insulator layer with magnetic 
impurities.~\cite{Bulaevskii,Spivak} More recently, it has been
proposed that $\pi$ junctions can be made by 
unconventional superconductors with non-$s$-wave symmetry.
Geshkenbein {\it et al.}~\cite{Geshkenbein} have discussed
$\pi$ junctions formed in heavy electron systems, whereas
$\pi$ junctions in high-$T_{\rm c}$ superconductors have been 
proposed to explain the positive
paramagnetic Meissner effect.~\cite{Braunisch,Sigrist}
In order to probe the symmetry of the superconducting gap,
a number of experiments have been performed,
including interference measurements
in single crystal $\mbox{YBa}_2\mbox{Cu}_3\mbox{O}_{7-\delta}$-Pb
SQUIDs,~\cite{Wollman,Brawner}
and direct imaging of magnetic flux in a tricrystal ring 
geometry.~\cite{Tsuei,Kirtley}
These measurements indicate that $\pi$ junctions can be realized at a
grain boundary of high-$T_c$ superconductors, and that
the $d$-wave symmetry is realized in high-$T_{\rm c}$ superconductors.

Static behavior of LJJs with both 0 and $\pi$ junctions
(0-$\pi$ junctions) has been studied 
theoretically by several authors.~\cite{Bulaevskii,Chen,Xu}
They have shown that a half vortex 
appears spontaneously 
at a boundary between 0 junction and $\pi$ junction
in sufficiently long junctions. 
Further, Kuklov~{\it et al.} have proposed that 
the half vortex can change its orientation 
by applying an external current 
to the junction.~\cite{Kuklov} They have also conjectured that
the half-vortex can be utilized in superconducting memory and
logic devise.

In this paper, we consider a 0-$\pi$-0 long Josephson junction,
which has a positive critical current at $|x|>a$, and has
a negative critical current at $|x|<a$. 
This junction may be realized
in systems such as grain boundary Josephson junctions.~\cite{Kirtley}
First, we study static properties of the junction
in $\S$~\ref{sec2}.
After that, we consider MQT in this junction 
to study quantum effects in $\S$~\ref{sec3}.
Summary is given in $\S$~\ref{sec4}.

\section{Hamiltonian and Static Properties}\label{sec2}
\subsection{Hamiltonian}
The Hamiltonian of the 0-$\pi$-0 Josephson junction
with an external current is given by
\begin{equation}
H = \int_{-\infty}^{\infty} \left( \frac12 \phi_x^2 +
\Theta(x) (1-\cos \phi) + f \phi \right) {\rm d}x.
\label{Ham}
\end{equation}
Here, $x$ is scaled by $\lambda_{\rm J}$,
and $f=I/I_0$ is an external current, 
and the Hamiltonian is scaled by an energy scale of the
junction, $E_0$. $\Theta(x)$ is a step function defined by
\begin{equation}
  \Theta(x) = \left\{ \begin{array}{cl}
    -1 & ( |x| < a ) \\ 1 & ( |x| > a) \end{array} \right. .
\end{equation}
Here, to simplify the situation,
we assume that the absolute value of the critical current is
uniform along the junction.
Static behavior of $\phi$ is given by $\delta H/\delta \phi(x)
=0$. From (\ref{Ham}), we obtain the sine-Gordon equation~\cite{Owen}
\begin{equation}
\phi_{xx} = \Theta (x) \sin \phi + f.
\label{sineGordon}
\end{equation}
This equation is also obtained from (\ref{dampedsineGordon})
by taking $\phi_t=0$. It should be noted
that dissipation does not affect static properties. 

\begin{figure}[tb]
\hfil
\epsfile{file=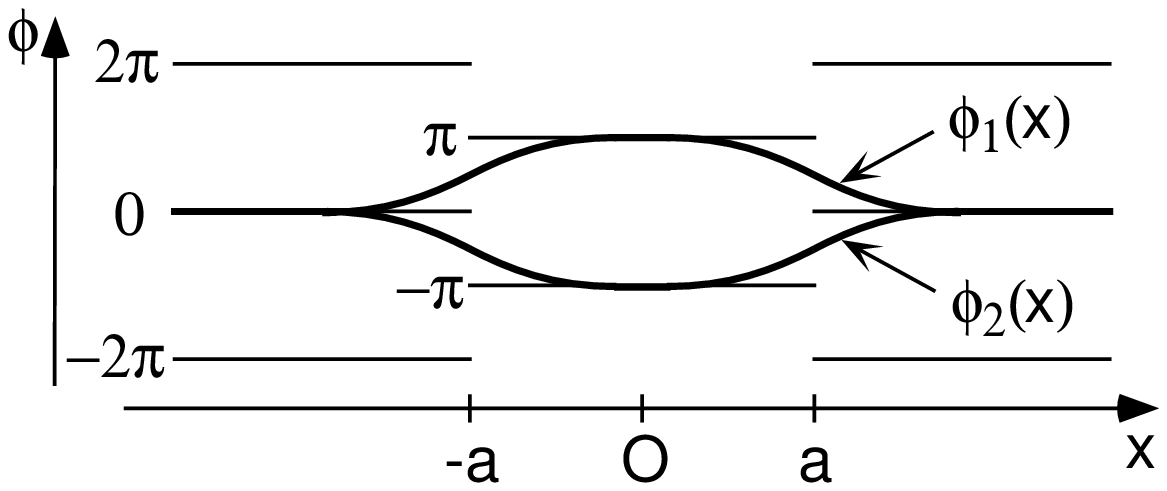,scale=0.65}
\hfil
\caption{Two types of solutions of (\ref{sineGordon}) are drawn
for a sufficiently large $a$.
The minima of the potential for $\phi$ is given as $(2n+1)\pi$ at
$|x|<a$, and as $n\pi$ at $|x|>a$.}
\label{staticsol}
\end{figure}

First, we consider the case with $f=0$.
In stable solutions by minimizing (\ref{Ham}), the system prefers
to have a uniform phase difference $\phi$ at $2\pi n$ for $x\gg|a|$, 
while $\phi=(2n+1)\pi$ is preferred for $|x|\ll a$, 
where $n$ is an integer.
As a result, we obtain two types of solutions, 
$\phi_1(x)$ and $\phi_2(x)$ as schematically 
shown in Fig.~\ref{staticsol}.
Here, we assumed without losing generality
that both solutions satisfy $\phi(\pm \infty) = 0$.

The physical meaning of these solutions is clear
in case of $a \gg 1$. For example, we consider 
the situation described by $\phi_1(x)$. 
The induced magnetic field in the junction 
is proportional to $\phi_x$. Hence, a vortex
and an antivortex appear at $x= -a$ and at $x=a$, respectively.
The magnetic flux carried by these vortex is $\pm\Phi_0 /2$,
where $\Phi_0=h/2e$ is the unit flux.
These induced vortices are called half vortices. 
The other solution $\phi_2(x)$ has the same vortices, except that
the orientations of the vortices are reversed
from those in $\phi_1(x)$.

\begin{figure}[tb]
\hfil
\epsfile{file=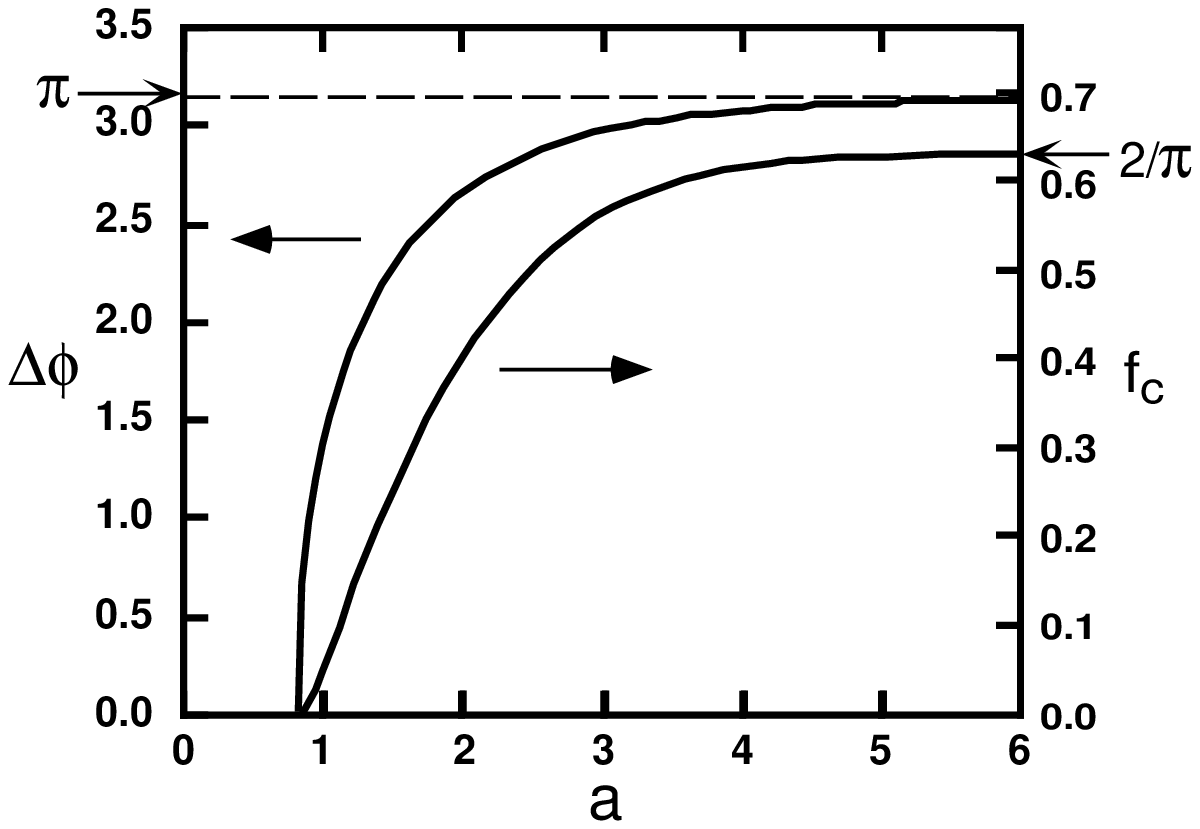,scale=0.7}
\hfil
\caption{$\Delta \phi = \phi_1(0) - \phi_1(\infty)$  
is shown as a function of $a$, which is a half of the distance
between two vortices scaled by $\lambda_{\rm J}$.
The critical current $f_{\rm c}$ 
necessary for vortices to change their orientation is also shown.
Both varnishes at $a=\pi/4\approx 0.79$ as $a$ decreases. }
\label{Deltaphi}
\end{figure}

\subsection{Magnetic flux}
The induced magnetic flux is proportional to $\Delta \phi=
\phi_1(0)-\phi_1(\infty)$. In the case of $a \gg 1$, 
we obtain $\Delta \phi = \pi$,
where the magnetic flux carried by the vortices is 
$\pm\Phi_0/2$. The magnetic flux defined by $\Delta \phi$, however, 
decreases as we put vortex and antivortex closer each other.
We have calculated $\phi(x)$ from (\ref{sineGordon})
numerically, and obtained $\Delta \phi$ as a function of $a$. 
We show the result in Fig.~\ref{Deltaphi}.
For $a \leq \pi /4 \approx 0.79$, we obtained 
$\Delta \phi = 0$. This means
that the vortices disappear when the distance
between vortices is too small. 
As shown later, the critical value $\pi/4$
can be obtained analytically. 
For $a \geq \pi /4$, the value of 
$\Delta \phi$ increases quickly as the value of $a$ 
becomes large, and for $a \gg 1$, approaches $\pi$. 

We expect that this behavior of the flux carried by
the vortices may be applied to accurate 
measurements of $\lambda_{\rm J}$. 
When we change the length of $\pi$ junction or
the Josephson penetration length $\lambda_{\rm J}$ by external magnetic
field, measurement of magnetic flux carried by vortices
will tell us the information about
the value of $\lambda_{\rm J}$. The measurement of
$\lambda_{\rm J}$ using direct imaging by scanning SQUIDs
has already been reported by Kirtley {\it et. al.}~\cite{Kirtley}
Compared with this, we expect that
the measurement using the 0-$\pi$-0 Josephson junction 
is more accurate, because the magnetic flux is sensitive
to the ratio $a=d/2\lambda_{\rm J}$, where $d$ is a length of
the $\pi$ junction.

\subsection{External current}
Next, we study static behavior of the solution in the presence of
the external current $f$.
We assume that the initial state is described by $\phi_1(x)$.
As the external current $f$ adiabatically
increases from zero, the form of $\phi_1(x)$ is modified.
The modified solution is denoted as $\phi_1(x,f)$.
When the external current takes a critical value $f_{\rm c}$, 
the solution $\phi_1(x,f)$ becomes unstable.
Then, a transition from $\phi_1(x,f)$ to the other
stable solution $\phi_2(x,f)$ occurs.
Here, $\phi_2(x,f)$ is a solution modified adiabatically from
the initial solution $\phi_2(x)$ by the external current $f$.
During the transition, a voltage pulse across the junction is generated.
This transition is intuitively understood easily as follows: 
the vortices exchange their locations each other. 
After that, the system remains
the state described by $\phi_2(x,f)$, as long as $f>0$.
When a negative external current is applied to the junction,
a transition from $\phi_2(x,f)$ to $\phi_1(x,f)$ occurs at $f=-f_{\rm c}$.

The critical current $f_{\rm c}$ is calculated from (\ref{sineGordon})
numerically. The result is shown also in Fig.~\ref{Deltaphi}.
For $a>\pi/4$, the critical current $f_{\rm c}$ increases
as $a$ becomes large, and is saturated toward $2/\pi$,
which can be obtained analytically as shown later.
Note that the critical value $f_{\rm c}$ is smaller than 1.
This means that the exchange of a vortex and an antivortex occurs
before the whole junction is driven to a voltage state.

\subsection{Limiting cases}
Next, we consider two limiting cases: (i) $a = \pi/4 + \lambda$
($\lambda \ll 1$), and (ii) $a \gg 1$. In these cases,
we can perform analytical calculation. 

In the case (i), the phase difference $\phi(x)$ satisfies 
$|\phi(x)|\ll 1$ for all $x$. 
Then, the Hamiltonian 
(\ref{Ham}) for $f=0$ up to $\phi^2$ is reduced to 
\begin{equation}
H[\phi(x)] = \int_{-\infty}^{\infty} \! {\rm d}x 
\, \vphi(x) \frac12 
\left( - \frac{{\rm d}^2}{{\rm d}x^2} + \Theta(x) \right) \vphi(x),
\label{appHam}
\end{equation}
where $\vphi$ represents fluctuations around the trivial solution
and is defined by $\phi(x) = 0 + \vphi(x)$. 
The eigenmodes of the fluctuation 
are obtained by solving the `Sch\"odinger equation'
\begin{equation}
-\frac{{\rm d}^2\vphi_n}{{\rm d}x^2} 
+ \Theta(x) \vphi_n = \eps_n \vphi_n,
\label{Scheq}
\end{equation}
under the normalization condition
\begin{equation}
\int_{-\infty}^{\infty} \! {\rm d}x 
\, \vphi_n(x) \vphi_m(x) = \delta_{nm} .
\label{normcond}
\end{equation} 
Here, $\Theta(x)$ can be regarded as a well-shaped potential 
with its width $2a$.  When $a \leq \pi/4$, the lowest energy
$\eps_0$ is positive, and the trivial solution $\phi=0$ 
is stable. However, when
$a = \pi/4 +\lambda$ ($0 < \lambda \ll 1$), there exists 
a negative eigenmode $\phi_0$, which means the instability
of the trivial solution $\phi(x)=0$.
Hence, when we expand the static solution
\begin{equation}
\phi(x) = \sum_{n=0}^{\infty} C_n \vphi_n(x),
\end{equation}
the coefficients $C_n$ become 0 for $n \geq 1$, and 
only $C_0$ is nonzero. For $a=\pi/4 + \lambda \: (\lambda \ll 1)$,
the ground state energy $\eps_0$ is close to zero. 
By taking $\eps_0 = 0$,
the form of the unstable mode $\vphi_0(x)$
is obtained approximately from (\ref{Scheq}) and (\ref{normcond}) as 
\begin{equation}
\vphi_0(x) = \left\{ \begin{array}{ll} 
\displaystyle{\sqrt{\frac{4}{\pi + 4}} \cos x } 
& (|x| <a), \\
\displaystyle{\sqrt{\frac{4}{\pi + 4}} \cos a \: e^{-(|x|-a)} } 
& (|x| >a). 
\end{array} \right. 
\label{vp0}
\end{equation}

To determine $C_0$, we derive
an effective Hamiltonian for $C_0$ by substituting
$\phi(x) = C_0 \vphi_0(x)$ and (\ref{vp0}) 
to the Hamiltonian (\ref{Ham}). By assuming that $C_0$ is small, 
a simple analysis gives 
\begin{equation}
H = - \frac{\eps_0}{2} C_0^2 
+ \frac{\pi+2}{8(\pi+4)^2} C_0^4 
- \sqrt{\frac{32}{\pi+4}} f C_0
\label{effHam1}
\end{equation}
to the fourth order of $C_0$. Here, $\eps_0=8\lambda/(\pi+4)$. 
From (\ref{effHam1}), $\Delta \phi$ and $f_{\rm c}$ are calculated 
analytically as
\begin{equation}
\Delta \phi = \sqrt{\frac{64}{\pi+4}} \lambda^{1/2}, 
\: \: f_{\rm c} = \frac{128}{27(\pi+2)} \lambda^{3/2}.
\end{equation}

In the case (ii), i.e. for $a \gg 1$, 
we can regard the junction as two independent
0-$\pi$ junctions. Then, a half vortex
and a half antivortex appear at $x=\pm a$. 
We notice only the half vortex at $x=a$.
The behavior of a half vortex in 0-$\pi$ junctions in the presence of 
an external current has already been studied by Kuklov 
{\it et.al.}~\cite{Kuklov} They have 
studied a critical current $f_{\rm c}$,
at which a half vortex changes its orientation with  
an integer flux being created.
In order to study the stability of the static solution $\phi_1(x,f)$,
we expand $\phi(x)$ around $\phi_1(x,f)$
\begin{equation}
\phi(x) = \phi_1(x,f) + \sum_{n=0}^{\infty} C'_n\vphi_n(x)
\end{equation}
with eigenmodes of small fluctuation defined by
\begin{equation}
-\frac{{\rm d}^2\vphi_n}{{\rm d}x^2} + U'' \vphi_n = E_n \vphi_n, 
\: \: U'' \equiv \frac{\partial^2 U}{\partial \phi^2}
(\phi_1(x,f)) .
\label{modeeq}
\end{equation}
Here, $U(\phi) = \Theta(x) (1-\cos \phi)$.
At $f=f_{\rm c}$, the lowest energy $E_0$ becomes zero as
in the case (i). It can easily be checked that
$\vphi_0 = C \partial_x \phi_1(x,f)$ always satisfies 
the equation (\ref{modeeq}) with $E_0=0$.
Here, the constant $C$ is determined by solving the normalization
condition (\ref{normcond}).
This mode is called the translational mode, because
this modulation translates the half vortex along the junction.
We should note, however, that both $\vphi_0(x)$ and $\partial_x
\vphi_0(x)$ must be continuous at $x=a$. This condition
and (\ref{sineGordon}) leads to $\phi_1(a,f_{\rm c})=0$.
Further, from the first integral of (\ref{sineGordon}), we obtain
\begin{eqnarray}
 0 - \cos \phi(0) -f\phi(0)
&=& \frac12 \phi_x(a)^2 - \cos \phi(a) -f\phi(a), \\    
 0 + \cos \phi(\infty) -f\phi(\infty) 
&=& \frac12 \phi_x(a)^2 + \cos \phi(a) -f\phi(a).     
\end{eqnarray}
From these equations and $\phi_1(a,f_{\rm c})=0$, 
the critical current for $a\gg 1$
is calculated as $f_{\rm c} = 2/\pi$.~\cite{Kukloverr}
For the critical solution $\phi(x,f_{\rm c})$, the normalization constant
in $\vphi_0$ is numerically calculated as $C\approx 0.5468$.

As well as in the case (i), only the lowest energy mode
$\vphi_0(x)$ becomes unstable for $f>f_{\rm c}$. 
The effective Hamiltonian in the case (ii) can be made
by substituting $\phi=\phi_1(x,f_{\rm c}) + C'_0 \vphi_0(x)$
to (\ref{Ham}) as 
\begin{equation}
H = \pi \delta C C'_0 - \alpha (C C'_0)^3,
\label{effHam2}
\end{equation}
to the third order of $C'_0$.
Here, $\delta = f_{\rm c} - f$ denotes the difference between
the critical current and the external current, and $\alpha$ 
is a constant calculated by
\begin{eqnarray}
\alpha &=& \frac16 \int_{-\infty}^{\infty} {\rm d}x \: \Theta(x) 
\sin \phi_1(x,f_{\rm c}) (\partial \phi_1(x,f_{\rm c}))^3  \nonumber \\ 
&=& \frac23 (\phi_{\infty} \sin \phi_{\infty} + 
\cos \phi_{\infty} -1) \approx 0.1403 ,
\end{eqnarray}
where $\phi_{\infty} = \phi(\infty) = -\sin^{-1} f_{\rm c}$.
From the effective Hamiltonian (\ref{effHam2}), 
it is shown that 
there exists a metastable state for $\delta > 0$.
Note that (\ref{effHam2}) is valid only for a small $C'_0$.
Hence, (\ref{effHam2}) can be used only for $\delta \ll 1$ case,
where the value of $C'_0$ at a metastable state is small.

\begin{figure}[tb]
\hfil
\epsfile{file=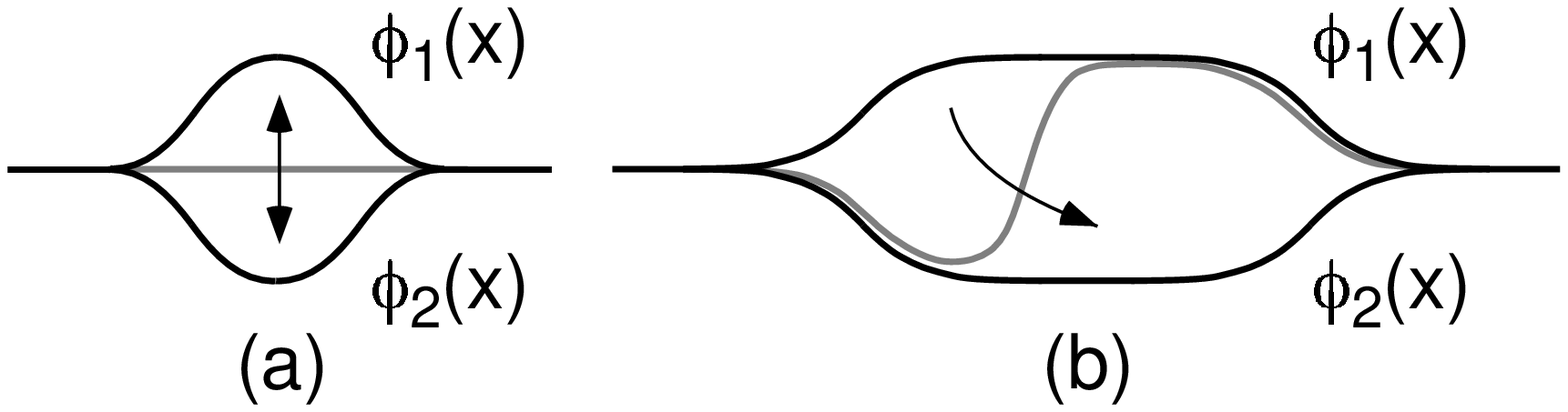,scale=0.4}
\hfil
\caption{Sketches of the transition from $\phi_1(x)$
to $\phi_2(x)$ in two limiting cases: (a)
$a=\pi/4 + \lambda \: (\lambda \ll 1)$, and
(b) for $a \gg 1$. 
The gray lines represent intermediate configurations
during the transition.
The transition occurs symmetrically
in the case (a), while the integer flux is generated 
at either side in the case (b).}
\label{exchange}
\end{figure}

\subsection{Crossover region}
It should be noted that  
the way to exchange the location of a vortex and an antivortex 
each other is different in the two limiting cases.
In the case (i), the transition from $\phi_1$ to $\phi_2$
occurs by keeping the symmetry $\phi(-x) = \phi(x)$
as shown in Fig.~\ref{exchange}(a),
because the most unstable mode $\vphi_0(x)$ is symmetric.
In the case (ii), the transition occurs symmetrically
in the original model (\ref{Ham}) as well as in the case (i). 
However, in the presence of even small spatial inhomogeneities,
either of two half vortices begins to move 
at a lower external current as shown in Fig.~\ref{exchange}(b).
Then, this half vortex changes its orientation first 
and creates an integer vortex in the region $|x|<a$.
This integer flux propagates along the $\pi$ junction toward the other half
vortex, and combines to the other half vortex to change its orientation. 

\begin{figure}[tb]
\hfil
\epsfile{file=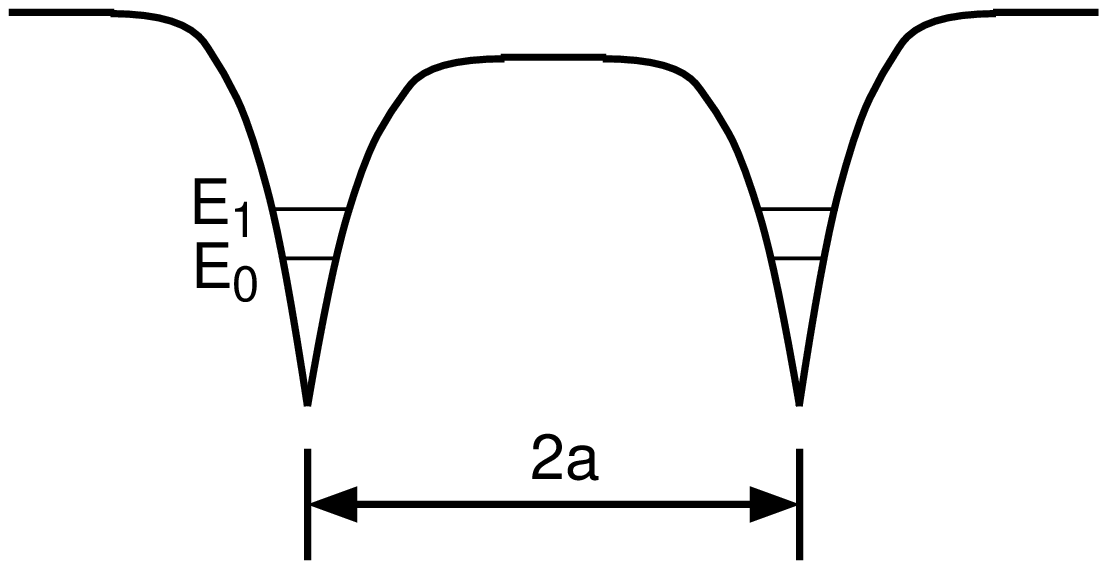,scale=0.5}
\hfil
\caption{A sketch of the potential $U''(\phi_1(x))$ in eq.~(\ref{Sch2}).
Here, $E_0$ is the ground state energy, and $E_1$ is the first excited state 
energy. }
\label{pot}
\end{figure}

To study the transition process for an intermediate value of $a$,
we focus on the Schr\"odinger equation for fluctuations $\vphi(x)$,
\begin{equation}
-\frac{{\rm d}^2\vphi_n}{{\rm d}x^2} + U'' \vphi_n = E_n \vphi_n, 
\: \: U'' \equiv \frac{\partial^2 U}{\partial \phi^2}
(\phi(x)) .
\label{Sch2}
\end{equation}
The potential term $U''(\phi(x))$ has two minima at $x = \pm a$
as shown in Fig.~\ref{pot}. 
The ground state energy $E_0$ and the first excited state
energy $E_1$ are also drawn in Fig.~\ref{pot}. As $a$ increases,
the wave function of the ground state is modified
by keeping the symmetry $\vphi_0(-x)=\vphi_0(x)$ and 
its nodeless form.
Hence, the most unstable mode $\vphi_0(x)$ is connected 
from the case (i) to the case (ii), smoothly.

When the distance between two wells $2a$
increases, the energy splitting $\Delta = E_1-E_0$
is suppressed exponentially.
For $a \gg 1$, the lowest two eigenstates become almost degenerate, 
and we can constitute wave functions localized at each well as
\begin{equation}
  \vphi_{R(L)} = (\vphi_0(x) \pm \vphi_1(x))/\sqrt{2}.
\end{equation}
In the presence of spatial inhomogeneities,
the potential energy is modified, and
the energy difference between two wells appears.
This effect can be studied by the effective Hamiltonian
on the two-dimensional Hilbert space spanned by $\vphi_R$ and
$\vphi_L$:
\begin{equation}
H_{\rm eff} = \Delta \sigma_x + \eps \sigma_z .
\end{equation}
Here, $\eps$ is the energy difference between wells due to
inhomogeneities, and $\sigma_x$ and $\sigma_z$ are Pauli's
matrices. When the energy splitting $\Delta$
is much larger than $\eps$, 
the wave function of the ground state is symmetric,
and given by $\vphi_0$ approximately.
Hence, the transition from $\phi_1$ to $\phi_2$ occurs
symmetrically as shown in Fig.~\ref{exchange}(a). 
On the other hand, $\Delta$ becomes
smaller than $\eps$ when $a$ is large. 
Then, the wave function of the ground state is localized in either
well, and given by $\vphi_R$ (or $\vphi_L$) approximately.
As a result, the transition begins from either side 
as shown in Fig.~\ref{exchange}(b).
The qualitative change which is expected to occur in the 
intermediate region of $a$ is not a transition but a crossover.

\begin{figure}[tb]
\hfil
\epsfile{file=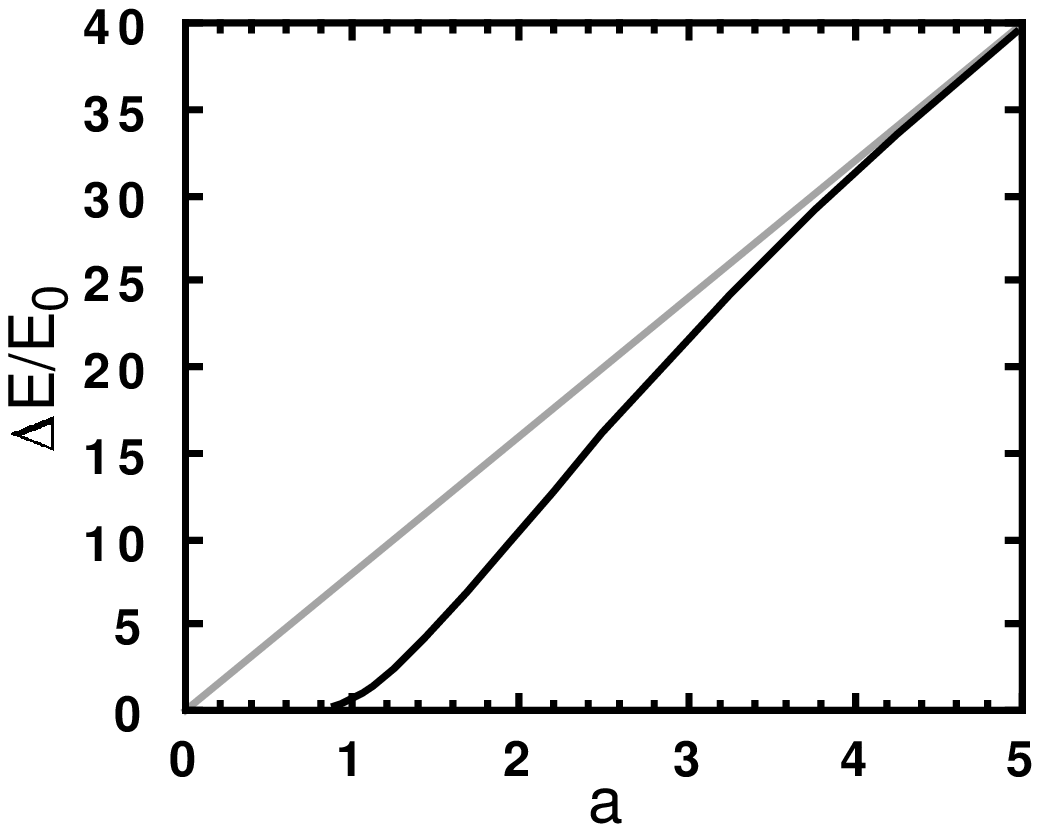,scale=0.7}
\hfil
\caption{The solid line represents
$\Delta E = H[\phi_1(x,f_{\rm c})]-H[\phi_2(x,f_{\rm c})]$
as a function of $a$. Here, $E_0$ is an energy scale of the junction,
and is defined below (\ref{deltaene}). The curve for $\Delta E
/ E_0$ approaches $8a$(gray line) for $a\gg 1$. }
\label{deltae}
\end{figure}

\subsection{Voltage pulse}
In order to estimate a voltage of the pulse,
we have
calculated numerically the energy difference $\Delta E$ defined as
\begin{equation} 
\Delta E/E_0 = H[\phi_1(x,f_{\rm c})] - H[\phi_2(x,f_{\rm c})] , 
\label{deltaene}
\end{equation}
where $E_0 = \Phi_0 I_0 \lambda_{\rm J}/2\pi L$ is
an energy scale of the junction, and $L$ is the junction length.
The result is shown in Fig.~\ref{deltae}. 
The energy difference increases monotonically for $a \geq \pi/4$.
When $a$ is large, the curve for $\Delta E/E_0$ approaches $8a$.
This feature can be obtained analytically by neglecting 
the spatial distribution of vortices around the boundary
between the 0 junction and the $\pi$ junction.
The electric power of the pulse $P$ is estimated as 
$P = \Delta E/T$, where $T$ is the time scale of the transition
from $\phi_1$ to $\phi_2$ or {\it vice versa}.
Accurate estimate of $T$ is difficult, because
we have to solve the sine-Gordon equation (\ref{dampedsineGordon})
dynamically. For $a \gg 1$, however, it is inferred 
that $T$ is determined
by the time for an integer vortex to propagate from $x=a$ to $x=-a$
or {\it vice versa}, and is estimated as $T\sim2a\lambda_{\rm J}/c$
in the original unit. Here, $c = \lambda_{\rm J} \omega_{\rm c}$ 
is a characteristic velocity of the integer vortex,
and $\omega_{\rm c} \sim \omega_{\rm p} \times \max(1,\alpha+\beta/3)$ 
is a characteristic frequency of the junction.
Thus, the pulse voltage $V$ is estimated as
\begin{equation} 
V \sim \frac{P}{I} \sim \frac{\Phi_0 \omega_{\rm c} \lambda_{\rm J}} 
{L}, 
\label{est}
\end{equation}
where $L$ is the junction length. Here, we used the critical current
$f_c = I/I_0= 2/\pi$. From Fig.~\ref{deltae},
it is expected that the pulse voltage is suppressed 
as $a$ decreases. Hence, the estimate (\ref{est}) is expected to 
give an upper limit for $V$.

\section{Macroscopic Quantum Tunneling}\label{sec3}
In this section, we consider 
the transition due to macroscopic quantum tunneling (MQT) 
from the metastable state $\phi_1$ to the stable state $\phi_2$.
In this paper, we neglect dissipation effects, taking
$\alpha$ and $\beta$ in (\ref{dampedsineGordon}) as 0.
Moreover, we study only limiting cases, which allow us
to perform analytical calculation on the basis of the
effective Hamiltonian (\ref{effHam1}) and (\ref{effHam2}).

We first consider the case  
$a = \pi/4 + \lambda \: (\lambda \ll 1)$.
We assume that $\delta = f_{\rm c} - f \ll 1$, and that
(\ref{effHam1}) can be approximated as a cubic potential. 
For a cubic potential,
the tunneling rate $\Gamma$ is calculated as~\cite{Weiss}
\begin{equation}
\Gamma = A \exp( -B), \: \: \: B=\frac{7.2V}{\hbar \omega_0}.
\label{rate}
\end{equation}
Here, $V$ is the energy barrier, and 
$\omega_0$ is the frequency of small oscillations
around the metastable state. The prefactor $A$ is the
order of $\omega_{\rm p}$, where $\omega_{\rm p}$ is the plasma frequency
of the junction. From (\ref{effHam1}), we can estimate the exponent
\begin{equation}
B \approx \frac{121}{\beta^2}\lambda^{-3/8}\delta^{5/4},
\label{exponent1}
\end{equation} 
where $\beta^2 = \hbar \omega_{\rm p} / E_0$.

In a similar way, the exponent $B$ is estimated 
for the case $a\gg 1$ from the effective Hamiltonian 
(\ref{effHam2}) as
\begin{equation}
B \approx \frac{99}{\beta^2}\delta^{5/4} .
\label{exponent2}
\end{equation}
Note that the exponent $B$ is proportional to $\delta^{5/4}$
in both cases as seen in (\ref{exponent1}) and (\ref{exponent2}).
Therefore, we expect that the $\delta$-dependence in $B$
does not change qualitatively in all the ranges of $a$.

The value of $\beta^2$ can be related to experimental parameters as
\begin{equation}
  \beta^2 = \frac{16\pi}{137}\left(\frac{2\lambda_L d}{W^2 \eps_{\rm r}}
  \right)^{1/2} ,
\end{equation}
where $\lambda_L$ is the London length, and $\eps_{\rm r}/d$ is
the capacitance per area, and $W$ is the junction width.
The typical experimental value for $\beta$
is very small ($\sim 10^{-3}$).~\cite{Kato} Therefore, MQT can be 
observed only for $\delta \ll 1$, because the tunneling rate
$\Gamma$ must be large enough to be observed in the laboratory.
For example, assuming $a \gg 1$, $\beta=10^{-3}$
and $A = 10^{10} {\rm [1/s]}$, we obtain $\Gamma \sim 2\times 10^2
{\rm [1/s]}$ for $\delta = 10^{-3}$.

The quantum effects
were treated here on the basis of the effective Hamiltonians
including only one degree of freedom for the field.
Other degrees of freedom appears only in the form of plasmons
in the semiclassical approximation, while
the plasmons do not affect the tunneling rate
at sufficiently low temperatures.~\cite{Kato}
The semiclassical approximation is
justified for $\beta \ll 1$. Then, many-body effects
characteristic of the sine-Gordon fields do not appear
in this junction.

In the above calculation, 
we have neglected dissipation effects due to quasiparticles.
It is, however, expected that damping effect
on the junction remains even at sufficiently low temperatures,
and strongly affects the tunneling rate,
when there exist gapless nodes for quasiparticle excitation 
in non-$s$-wave superconductors.
Because dissipation on the junction 
generally suppresses the tunneling rate, our calculation gives
an upper limit for $\Gamma$.~\cite{Caldeira} 
Damping effects on MQT in this case with more accurate 
estimate of the tunneling rate remains for further studies.

\section{Summary}
\label{sec4}
In summary, we studied static properties of magnetic fluxes 
and their macroscopic quantum tunneling in the 0-$\pi$-0 Josephson
junction, where two half vortices are formed if the $\pi$ junction
region is long.
We calculated the magnetic flux of spontaneously
induced vortices, and the critical current needed to make
a transition between two degenerate vortex configurations.
We also studied quantum tunneling rate for this transition.
This MQT may be observed in high-$T_{\rm c}$ superconductors 
under an appropriate condition.

\end{document}